\documentclass[aps,pra,showpacs,twocolumn,superscriptaddress]{revtex4}
\usepackage{amssymb}
\usepackage{amsmath}
\usepackage{graphicx}
\usepackage{dcolumn}
\usepackage{bm}

\begin{document}

\title{Bose-Einstein condensation in an optical lattice: A perturbation
approach}
\author{C. Trallero-Giner}
\affiliation{Faculty of Physics, Havana University, 10400 Havana, Cuba}
\author{Victor Lopez-Richard}
\affiliation{Departamento de Fisica, Universidade Federal de S\~{a}o Carlos, 13.565-905, S%
\~{a}o Carlos, S\~{a}o Paulo, Brazil}
\author{Ming-Chiang Chung}
\affiliation{Institute of Physics, Academia Sinica, Taipei 11529, Taiwan}
\author{Andreas Buchleitner}
\affiliation{Physikalisches Institut der Albert-Ludwigs-Universit\"at,
Hermann-Herder-Str. 3, D-79104 Freiburg}
\date{\today }

\begin{abstract}
We derive closed analytical expressions for the order parameter $\Phi (x)$
and for the chemical potential $\mu $ of a Bose-Einstein Condensate loaded
into a harmonically confined, one dimensional optical lattice, for
sufficiently weak, repulsive or attractive interaction, and not too strong
laser intensities. Our results are compared with exact numerical
calculations in order to map out the range of validity of the perturbative
analytical approach. We identify parameter values where the optical lattice
compensates the interaction-induced nonlinearity, such that the condensate
ground state coincides with a simple, single particle harmonic oscillator
wave function.
\end{abstract}

\maketitle

\section{Introduction}

The dynamics of Bose Einstein Condensates (BEC) in harmonically confined
optical lattices is one of the key objects of cold matter research \cite%
{shin}. While the fundamental equation which describes the mean field
solution, the Gross-Pitaevski equation (GPE) \cite{Groos1}, is readily
amenable to a numerical solution, explicit analytical expressions for the
condensate ground state $\Phi $ and the associated chemical potential $\mu $
are scarce. Variational solutions have been proposed in \cite{perez}, and
quasiclassical approximations were implemented, providing several analytical
results for the GPE \cite{Konotop}. Eigenenergies and eigenmodes of the
non-linear Schr\"{o}dinger equation in a parabolically confined optical
lattice can be characterized through asymptotic expansions of Mathieu
functions, in the tight-binding approximation \cite{ana}.

We will demonstrate in the present contribution that closed analytic results
can indeed be derived, and that they permit not only qualitative but also
quantitative insight into the competition between the interaction-induced
nonlinearity on one hand, and the redistribution of the particles over the
spatially modulated potential, on the other. Comparison with exact numerical
results will finally allow us to demarcate the range of validity of our
perturbation approach, with respect to parameter values employed in state of
the art experiments.

\section{Analytical expressions}

We consider the one-dimensional nonlinear GPE in a stationary optical
lattice \cite{Groos1},
\begin{eqnarray}
&&\left\{ -\frac{\hbar ^{2}}{2m}\frac{d^{2}}{dx^{2}}+\frac{1}{2}m\omega
^{2}x^{2}+\lambda _{1D}\left\vert \Phi \right\vert ^{2}\right.  \notag \\
&&\left. -sE_{R}\cos ^{2}\left( \frac{2\pi }{d}x\right) \right\} \Phi =\mu
\Phi ,  \label{GPOL}
\end{eqnarray}%
where $\omega $ is the frequency of the confining harmonic trap \cite{Exp1},
$m$ the atomic mass, $E_{R}=$ $\frac{\hbar ^{2}}{2m}(\frac{2\pi }{d})^{2}$
the recoil energy \cite{Kramer}, $s\geq 0$ an adjustable parameter defined
by the laser intensity, $d$ the laser wavelength, and $\lambda _{1D}$ the
atom-atom interaction strength. With this one dimensional ansatz, we
implicitly assume that the dynamics in the transverse direction be frozen --
e.g., by sufficiently strong confinement or some other adiabaticity
condition \cite{Berg}. Furthermore, we imply the number of atoms per site to
be large enough to justify the mean field limit \cite{Pitaevskii,Pitaevskii2}%
. The normalization condition for $\Phi $ reads
\begin{equation}
1=\int_{-\infty }^{\infty }dx\left\vert \Phi \right\vert ^{2}\,.
\end{equation}

Equation (\ref{GPOL}) admits the following integral representation for $\Phi
(x)$ \cite{trallero}:
\begin{equation}
\Phi (x)=\int_{-\infty }^{\infty }G(x,x^{\prime })f(x^{\prime })dx^{\prime
}\,,  \label{in}
\end{equation}%
with the Green function $G(x,x^{\prime })$ of the linear operator $L_{0}=-%
\frac{\hbar ^{2}}{2m}d^{2}/dx^{2}+\frac{1}{2}m\omega ^{2}x^{2}$, and%
\begin{equation*}
f=\left( \mu +sE_{R}\cos ^{2}\left( 2\pi x/d\right) -\lambda _{1D}\left\vert
\Phi \right\vert ^{2}\right) \Phi
\end{equation*}%
the inhomogeneity of the differential equation $L_{0}[\Phi ]=f$.

Recall that for Eq.~(\ref{in}) to hold, we have to impose that all involved
functions tend to zero as $x\rightarrow \pm \infty $. Thus, on the basis of
the general theory of Fredholm integral equations \cite%
{Mihling,petrovskii,trallero2}, the harmonic oscillator functions $\{\varphi
_{n}\}$ represent a complete basis set which span the Hilbert space of the
integral equation\ (\ref{in}). Using the spectral representation of $%
G(x,x^{\prime })$, we can rewrite the order parameter $\Phi $ as%
\begin{equation}
\Phi =\sum_{n=0}^{\infty }\varphi _{n}(\frac{x}{l_{o}})C_{n}(\mu )\,,
\label{fi}
\end{equation}%
where $l_{o}=\sqrt{\hbar /m\omega }$ is the harmonic oscillator wavelength.
Inserting (\ref{fi}) into (\ref{in}), the vector coefficient $\mathbf{C}(\mu
)=\{C_{n}(\mu )\}_{n=0}^{\infty }$ is restricted to obey the relation \cite%
{Mikhlin}
\begin{equation}
\left[ \mathbf{\Delta }(\mu )+\Lambda \mathbf{C}^{\tau }\mathbf{\cdot T}%
\cdot \mathbf{C-}V_{o}\mathbf{P}\right] \mathbf{C}=0\,,  \label{Hill}
\end{equation}%
with $\mathbf{C}^{\tau }$ the transpose of the vector $\mathbf{C}$. $T_{plmn}
$ and $P_{km}$ are fourth and second rank tensors, respectively, and defined
in Appendices A and B. Furthermore, $V_{o}=sE_{R}/\hbar \omega $, $\Lambda
=\lambda _{1D}/l_{o}\hbar \omega $, and
\begin{equation}
\mathbf{\Delta }_{nm}=\left( n+\frac{1}{2}-\frac{\mu }{\hbar \omega }\right)
\delta _{nm}\,.  \label{matriele}
\end{equation}

\subsection{Chemical potential}

We tackle the problem under the assumption that the atom-atom interaction,
the nonlinear term $\lambda \left\vert \Phi \right\vert ^{2}$, and the
optical lattice potential can be considered as perturbations with respect to
the trap potential $m\omega ^{2}x^{2}/2$. In this case, the dimensionless
chemical potential $\mu /\hbar \omega $ and the vector $\mathbf{C}$ in Eq.~(%
\ref{Hill}) can be expanded in series,
\begin{eqnarray}
\mathbf{C} &=&\mathbf{C}^{(0)}+\mathbf{C}^{(1)}+\mathbf{C}^{(2)}+\ldots \,,
\label{C} \\
\frac{\mu }{\hbar \omega } &=&\mu ^{(0)}+\mu ^{(1)}+\mu ^{(2)}+\ldots \,,
\label{mu}
\end{eqnarray}%
where the quantities $\mathbf{C}^{(i)}$ and $\mu ^{(i)}$ are understood to
be of the same order in $\Lambda $ and $V_{o}$. Expansion of Eq.~(\ref{Hill}%
) to second order in $\Lambda $ and $V_{o}$ leads, after lengthy but
straightforward manipulation, the following expression for the system's
ground state energy:
\begin{eqnarray}
\frac{\mu }{\hbar \omega } &=&\frac{1}{2}+\Lambda
T_{0000}-V_{o}P_{00}-3\Lambda ^{2}\sum_{m=1}^{\infty }\frac{\left\vert
T_{0002m}\right\vert ^{2}}{2m}+  \notag \\
&&4\Lambda V_{o}\sum_{m=1}^{\infty }\frac{T_{0002m}P_{02m}}{2m}%
-V_{0}^{2}\sum_{m=1}^{\infty }\frac{\left\vert P_{02m}\right\vert ^{2}}{2m}%
\,.
\end{eqnarray}%
Using the properties of $T_{plmn}$ and $P_{kp}$ given in the Appendices A
and B we obtain
\begin{eqnarray}
\frac{\mu }{\hbar \omega } &=&\frac{1}{2}+\frac{\Lambda }{\sqrt{2\pi }}-%
\frac{V_{o}}{2}\left[ 1+\exp \left( -\alpha ^{2}\right) \right] +  \notag \\
&&\frac{\Lambda V_{o}}{\sqrt{2\pi }}\exp \left( -\alpha ^{2}\right)
\sum_{m=1}^{\infty }\frac{\alpha ^{2m}}{m2^{m}m!}-  \notag \\
&&\frac{3\Lambda ^{2}}{2\pi }\sum_{m=1}^{\infty }\frac{(2m-1)!}{%
2^{4m}(m!)^{2}}-  \notag \\
&&\frac{V_{o}^{2}}{4}\exp \left( -2\alpha ^{2}\right) \sum_{m=1}^{\infty }%
\frac{2^{2m}\alpha ^{4m}}{2m(2m)!}\,,  \label{pert}
\end{eqnarray}%
with $\alpha =2\pi l_{o}/d$ the ratio of trap to laser wavelength. Finally,
the series in (\ref{pert}) can be summed up with the help of Eqs.~(\ref{1}-%
\ref{4}) in Appendix C, and we obtain for the chemical potential, at second
order in $\Lambda $ and $V_{o}$:
\begin{eqnarray}
\frac{\mu }{\hbar \omega } &=&\frac{\Lambda }{\sqrt{2\pi }}+\frac{1-V_{o}}{2}%
-\frac{V_{o}}{2}\exp \left( -\alpha ^{2}\right)  \notag \\
&&+\frac{\Lambda V_{o}}{\sqrt{2\pi }}\exp \left( -\alpha ^{2}\right) \left\{
Ei(\frac{\alpha }{2})-\mathcal{C}-\ln \frac{\alpha }{2}\right\}  \notag \\
&&-0.033106\times \Lambda ^{2}  \notag \\
&&-\frac{V_{o}^{2}}{4}\exp \left( -2\alpha ^{2}\right) \left\{ Chi(2\alpha )-%
\mathcal{C}-\ln 2\alpha \right\} .  \label{pertmu}
\end{eqnarray}

\subsection{Order parameter}

Following the same procedure as above, $\mathbf{C}$ and thus (through (\ref%
{fi})) the normalized order parameter $\Phi $ can be expressed, at first
order in $\Lambda $ and $V_{o}$, as
\begin{equation}
\Phi =\varphi _{_{0}}(\frac{x}{l_{o}})-\sum_{m=1}^{\infty }\frac{\Lambda
T_{0002m}-V_{o}P_{02m}}{2m}\varphi _{2m}(\frac{x}{l_{o}})\,.  \label{fiper}
\end{equation}%
With the expressions derived in the Appendices A and B for the matrix
elements $T_{0002m}$ and $P_{02m}$, this can be rewritten as
\begin{eqnarray}
\Phi &=&\varphi _{_{0}}(\frac{x}{l_{o}})+\sum_{m=1}^{\infty }\frac{\left(
-1\right) ^{m+1}}{2m}\left\{ \frac{\Lambda \sqrt{\left( 2m\right) !}}{\sqrt{%
2\pi }2^{2m}m!}-\right.  \notag \\
&&\left. \frac{V_{o}2^{m-1}}{\sqrt{\left( 2m\right) !}}\alpha ^{2m}\exp
\left( -\alpha ^{2}\right) \right\} \varphi _{2m}(\frac{x}{l_{o}})\,.
\label{fif}
\end{eqnarray}%
Since, according to Eqs.~(\ref{F}-\ref{Gf}) of Appendix C,
\begin{equation*}
F\left( z,\alpha \right) =\frac{\exp \left( -\alpha ^{2}\right) }{\sqrt{\pi
^{1/2}}}\sum_{m=1}^{\infty }\frac{\left( -1\right) ^{m}2^{m-1}}{2m\sqrt{%
\left( 2m\right) !}}\alpha ^{2m}\varphi _{2m}(z)
\end{equation*}%
equals
\begin{eqnarray*}
F\left( z,\alpha \right) &=&\frac{1}{\sqrt{l_{o}\pi ^{1/2}}}\exp \left(
-\alpha ^{2}\right) \exp \left( -z^{2}/2\right) \times \\
&&\int\nolimits_{0}^{\alpha }\left[ \exp (y^{2})\cos \left( 2yz\right) -1%
\right] \frac{dy}{2y}\,,
\end{eqnarray*}%
and, moreover, with (\ref{f}-\ref{gf}),
\begin{equation*}
G\left( z\right) =\sum_{m=1}^{\infty }\frac{\left( -1\right) ^{m+1}\sqrt{%
(2m)!}}{2m2^{2m}\sqrt{2\pi }m!}\varphi _{2m}(z)
\end{equation*}%
can be condensed into
\begin{equation*}
G\left( z\right) =\frac{\exp \left( -z^{2}/2\right) }{\sqrt{2\pi }\sqrt{%
l_{o}\pi ^{1/2}}}\int\nolimits_{1}^{\sqrt{2}/2}\frac{\exp \left( -\frac{z^{2}%
}{y^{2}}\left( 1-y^{2}\right) \right) -y}{1-y^{2}}dy\,,
\end{equation*}%
we obtain the following closed expression for the order parameter:
\begin{equation}
\Phi =\varphi _{_{0}}(\frac{x}{l_{o}})+\frac{\Lambda }{\sqrt{l_{o}}}G\left(
\frac{x}{l_{o}}\right) +\frac{V_{o}}{\sqrt{l_{o}}}F\left( \frac{x}{l_{o}}%
,\alpha \right) \,.  \label{ExaFi}
\end{equation}

\subsection{Numerical Solution}

For a numerical solution of Eq.~(\ref{GPOL}), we subdivide the dimensionless
coordinate $z=x/l_{o}$, $z\in \lbrack -L;L]$, such that $z_{i}=(-L/2+(i-1))%
\delta $, $i=1,\ldots ,L+1$, with a grid size $\delta $. For $\delta \ll
d/l_{o}$, a three-point approximation with uniform spacing allows to rewrite
the second derivative and the differential equation (\ref{GPOL}) as
difference equations:
\begin{eqnarray}
&&\left. \left( \frac{z_{i}^{2}}{2}+\Lambda \left\vert \overline{\phi _{i}}%
\right\vert ^{2}-V_{o}\cos ^{2}\frac{{2\pi {l}_{o}z_{i}}}{d}+\frac{1}{\delta
^{2}}\right) \overline{\phi _{i}}\right.  \notag \\
&&\left. -\frac{\overline{\phi _{i-1}}}{2\delta ^{2}}-\frac{\overline{\phi
_{i+1}}}{2\delta ^{2}}=\frac{\mu }{\hbar \omega }\overline{\phi _{i}}\right.
\,,  \label{DFE}
\end{eqnarray}%
where $\overline{\phi _{i}}=\sqrt{{{l}_{o}}}\Phi (z_{i})$ is the local,
normalized order parameter which abides by the boundary conditions $%
\overline{\phi _{0}}=\overline{\phi _{L+2}}=0$.
\begin{figure}[tb]
\includegraphics[width=85mm]{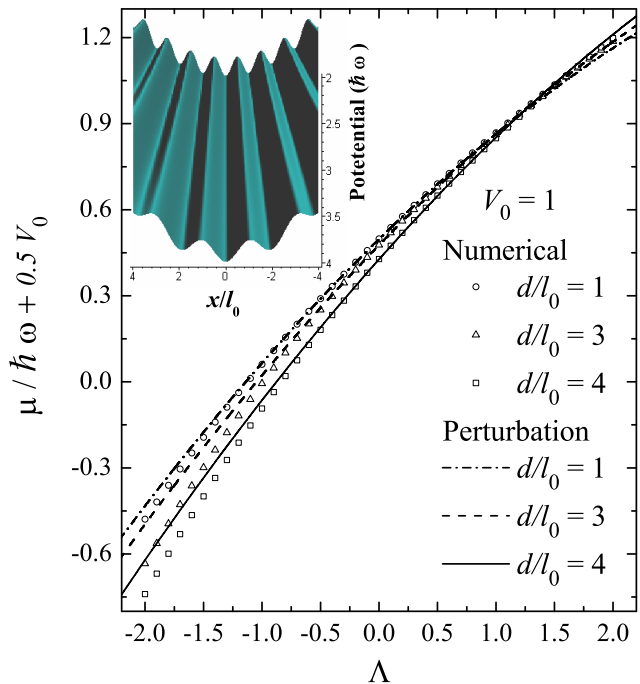}
\caption{(Color online) Variation of the chemical potential $\protect\mu %
/\hbar \protect\omega +V_{o}/2$ as a function of the dimensionless
self-interaction parameter $\Lambda $. Solid, dashed, and dash-dotted lines
represent the analytical results for $d/l_{o}=$ 1, 2 and 3, respectively.
Symbols correspond to the numerical solution of Eq.\ (\protect\ref{GPOL}).
The strength of the periodic potential was set to $V_{o}=sE_{R}/\hbar
\protect\omega =$ 1. Inset: Confining potential $U=\frac{1}{2}m\protect%
\omega ^{2}x^{2}-sE_{R}\cos ^{2}\left( \frac{2\protect\pi }{d}x\right) $ for
$2<d/l_{o}<4$.}
\label{fig1}
\end{figure}
Equation ({\ref{DFE})} defines a tri-diagonal eigenvalue problem with the
normalization condition $\sum_{i}|\overline{\phi _{i}}|^{2}\delta =1$ or,
equivalently, $\sum_{i}|\tilde{\phi}_{i}|^{2}=1$, where $\tilde{\phi}_{i}=%
\sqrt{\delta }\bar{\phi}_{i}$. Since the resulting matrix (\ref{DFE})
involves the non-linear term $\left\vert \overline{\phi _{i}}\right\vert
^{2} $, an iterative procedure is needed for a self-consistent solution. To
avoid numerical instabilities \cite{trallero2,Pu}, we inserted the iterative
wave function
\begin{equation}
F_{i}^{(k)}=\sqrt{(1-\varepsilon )\left\vert \tilde{\phi}_{i}^{(k-1)}\right%
\vert ^{2}+\varepsilon \left\vert \tilde{\phi}_{i}^{(k)}\right\vert ^{2}}
\end{equation}%
in (\ref{DFE}), at each new iteration $k$, with $\varepsilon \in \lbrack
0;1] $, and starting condition $F_{i}^{(0)}=\tilde{\phi}_{i}^{(0)}$. This
procedure was repeated until $\left\vert \tilde{\phi}_{i}^{(k)}-\tilde{\phi}%
_{i}^{(k-1)}\right\vert <\delta _{\phi }$ and $|(\mu ^{(k)}-\mu
^{(k-1)})/\hbar \omega |<\delta _{\mu }$, where $\delta _{\phi }$ and $%
\delta _{\mu }$ are the desired accuracies for the order parameter and the
chemical potential, respectively. Our iterative solutions then converge to a
relative uncertainty of $10^{-4}$ for $\mu /\hbar \omega $ and the
normalized order parameter.

\section{Results}

\begin{figure}[tb]
\includegraphics[width=80mm]{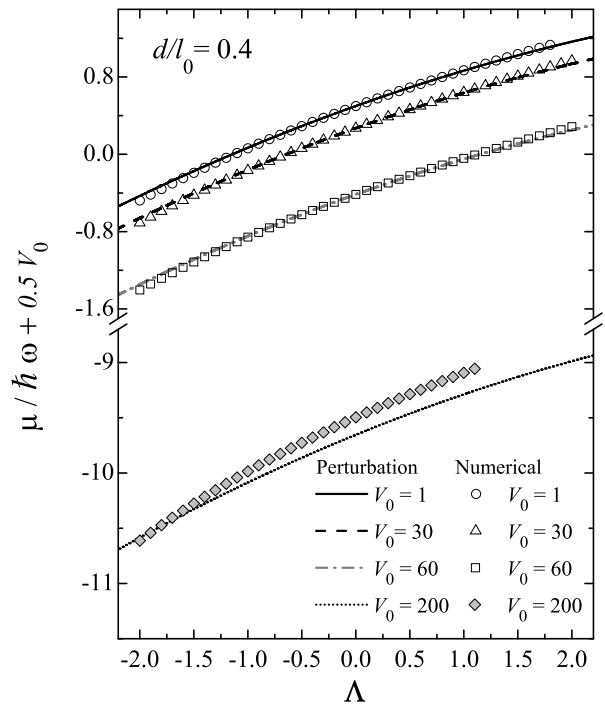}
\caption{Same as in Fig.\ \protect\ref{fig1}, for $V_o=$ 1 (solid line), 30
(dashed line), 60 (dash-dotted line), and 200 (dotted line). Symbols
represent the analytical results. The laser wavelength was set to $d/l_o=$%
0.4. }
\label{fig2}
\end{figure}

\begin{figure}[tb]
\includegraphics[width=85mm]{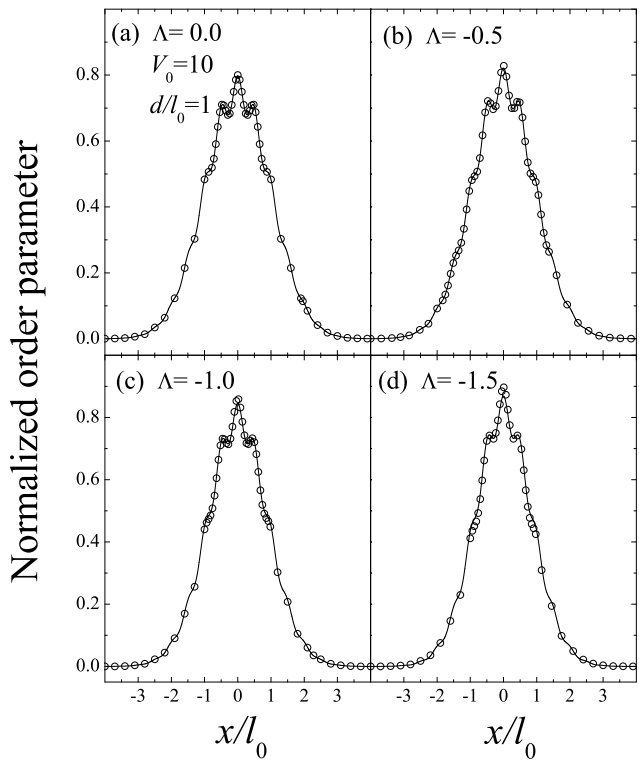}
\caption{Normalized order parameter $\protect\sqrt{l_{0}}\Phi (x/l_{0})$ for
the dimensionless self-interaction values: (a) $\Lambda =$0, (b) -0.5, (c)
-1.0, and (d) -1.5. $V_o=$ 1 and $d/l_o=$1. Solid line: Closed solution
according to Eq.\ (\protect\ref{ExaFi}). Open circles: Numerical result.}
\label{fig3}
\end{figure}

\begin{figure}[tb]
\caption{Order parameter $\Phi (x/l_{0})$ for $d/l_o=$ 1. (a) $\Lambda $ =2,
$V_{o}=sE_{R}/\hbar \protect\omega=$ 0.5, 1, 5, and 10. (b) $V_{o}=$ 10, $%
\Lambda $ =-2, 0, 2. Symbols represent the numerical solution. For the sake
of comparison, $U$ (see caption Fig.\ \protect\ref{fig1}) and the harmonic
oscillator potential are shown by solid and dashed lines, respectively.}
\label{fig4}\includegraphics[width=85mm]{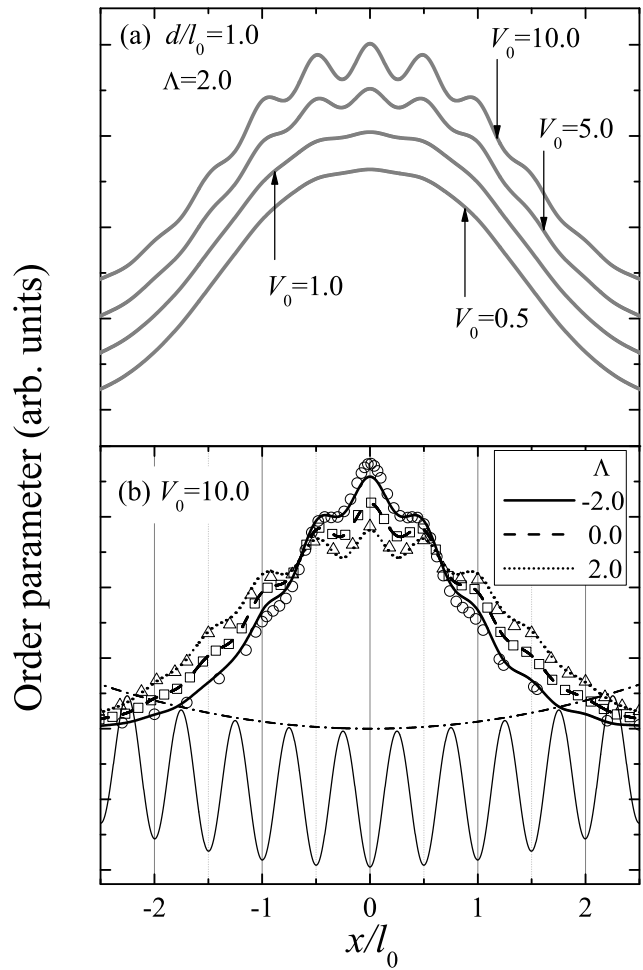}
\end{figure}

Figure\ \ref{fig1} displays the variation of the chemical potential, $\mu
/\hbar \omega +0.5V_{o}$, as a function of the dimensionless atom-atom
interaction $\Lambda $. The analytical solution (\ref{pert}) is represented
by solid, dashed, and dash-dotted lines, while symbols represent the
numerical solutions of Eq.~(\ref{GPOL}). The inset illustrates the confining
potential $U=m\omega ^{2}x^{2}/2-sE_{R}\cos ^{2}\left( 2\pi x/d\right) $, as
a function of $x/l_o,$ for $V_{o}=1$ and $2<d/l_o<4$. In our calculation of $%
\mu $, we fixed $V_{o}=1$ and checked the variation of the perturbative
solution as compared to the numerical one, for several values $d/l_{o}=1,3$
and $4$ of the normalized wavelength. For $\Lambda <0$ the relative error
between the numerical solution and Eq.~(\ref{pertmu}) increases as $d/l_{o}$
increases, while for $\Lambda >0$ the worst comparison is obtained at $%
d/l_{o}=1$. In the case of repulsive interaction, numerical and perturbative
calculation exhibit small variations with $d/l_{o}$. In general, Eq.~(\ref%
{pertmu}) describes the values of the chemical potential better for
repulsive than for attractive interaction. This can be understood with the
following argument: In \cite{trallero,trallero2} it was shown shown that, at
$V_{o}=0$, perturbation theory matches the numerical solution very well in
the interval $-2<\Lambda <2$. When the optical lattice is switched on ($%
V_{o}\neq 0$), we can argue that an effective renormalization of the
non-linear parameter takes place. Since Eq.~(\ref{pertmu}) includes a
negative quadratic term in $V_{o}$, we expect that the effective range of $%
\Lambda $ where perturbation theory is applicable will shift to $\Lambda >-2$%
, i.e., more repulsive or less attractive atom-atom interactions. Naturally,
the behavior of the chemical potential with $d/l_{o}$ limits this simple
description.

In Fig.~\ref{fig2}, we plot the dependence of $\mu (\Lambda )/\hbar \omega
+0.5V$ on $\Lambda $, for several values of the strength of the periodic
potential $V_{o}$. In the calculation we took $d/l_{o}=0.4$. The figure
shows perfect agreement of Eq.~(\ref{pertmu}) with the numerical result for $%
V_{o}$ ranging between $1$ and $60$, while, at $V_{o}=200$, we observe large
discrepancies.

\begin{figure}[tb]
\includegraphics[width=85mm]{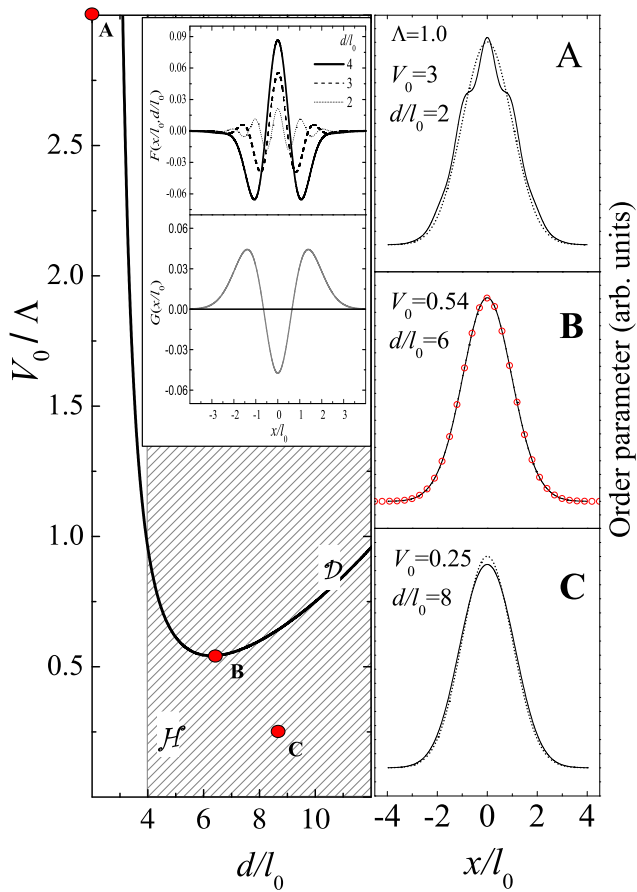}
\caption{(Color online) Quenching of the non-linear repulsive interaction.
Left panel: Region $\mathcal{H}$ is defined by the condition (\protect\ref%
{cond}). The curve $\mathcal{D}$ is defined by Eq.\ (\protect\ref{cond}), at
$x=0$. Right panel: Order parameter calculated for those parameter values
identified by \emph{A}, \emph{B}, and \emph{C} on the left. Solid lines
represent $\Phi$ as derived from Eq.\ (\protect\ref{ExaFi}), while open
circles indicate the exact numerical solution. Dashed lines correspond to
the harmonic oscillator wave function.}
\label{fig5}
\end{figure}

\begin{figure}[tb]
\includegraphics[width=85mm]{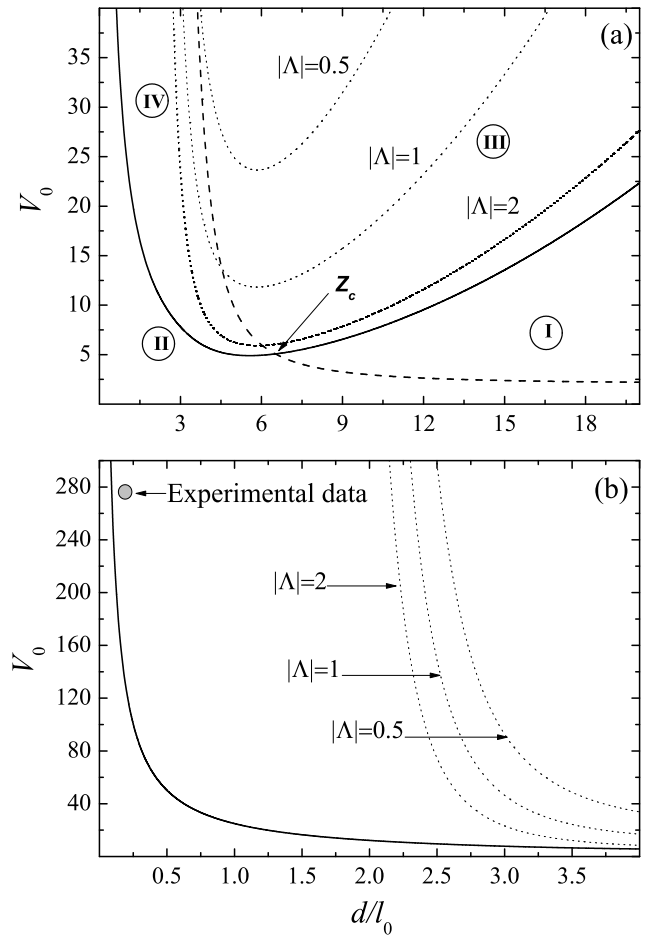}
\caption{ (a) Range of validity of Eq.\ (\protect\ref{pertmu}) in the $V_{o}$%
-$d/l_{o}$-plane. The functions \emph{Linear}, \emph{Mixed} and \emph{%
Quadratic} (see text) are represented by dashed, dotted and solid lines,
respectively. $z_{c}=[(V_{o})_{c},(d/lo)_{c}$ ] is defined by equality
between \emph{Linear} and \emph{Quadratic}. (b) Zoom into the region of
large values of $V_{o}$. The circle locates the parameter values employed in
the experiments reported in \protect\cite{Ott}.}
\label{fig6}
\end{figure}

Figure~\ref{fig3} displays the normalized order parameter, $\Phi (x/l_{0})%
\sqrt{l_{0}},$ for four values of $\Lambda =0,-0.5,-1.0$, and $-1.5$, and
fixed $V_{o}=10$, $d/l_{o}=1$. In general, we obtain a remarkably good
agreement between the analytical prediction (\ref{ExaFi}) and the numerical
result, for all considered (attractive) values of the dimensionless
interaction parameter $\Lambda $, and witness the interplay between the
optical lattice and harmonic oscillator potential: The harmonic oscillator
ground state $\varphi _{_{0}}$ is modulated by an oscillatory behavior,
induced by the optical lattice potential. We further note that the wave
function gets more localized, and the maximum of $\Phi (x/l_{0})$ increases,
as the non-linear potential becomes more attractive. Panel (a) of Fig.~\ref%
{fig4} shows the variation of $\Phi (x/l_{0})\sqrt{l_{0}}$ with the strength
of the periodic potential $V_{o}$, for the repulsive case $\Lambda =2$. As $%
V_{o}$ increases from 0.5 to 10, the lattice-induced density oscillations
manifest with increasing amplitude. A comparison between attractive and
repulsive interactions, for $V_{o}=10$, is given by panel (b) of Fig.~\ref%
{fig4}: Since $d/l_{o}=1$ in Figs.~\ref{fig3} and \ref{fig4}, the wave
function $\Phi (x/l_{0})$ exhibits oscillations at $x/l_{0}\approx n/2$ ($%
n=\pm 1,\pm 2,\ldots $), which are quenched because of the asymptotic decay $%
\propto $ $\exp \left( -x^{2}/2l_{o}^{2}\right) $ of the oscillator wave
functions. These oscillations are illustrated in Fig.~\ref{fig4}(b), where
the solid line represents the potential $U$ (see inset of Fig.~\ref{fig1}),
along with the wavefunction $\Phi $ for the same values of $V_{o}$ and $%
d/l_{o}$. The positions of the observed maxima (minima) in the order
parameter correspond to the minima (maxima) of the resulting effective
potential $U$.

The extrema of $\Phi $ exhibit different behavior depending on the
non-linear interaction $\Lambda $. As the condensate becomes effectively
less confined for increasing repulsive interaction, the amplitudes of the
maxima and minima of $\Phi $ are enhanced. From the physical point of view
it is clear that a strongly attractive interaction (in the figure: $\Lambda
=-2$) should flatten the lattice-induced oscillations of the condensate. In
panel (b) we compare the evolution of the order parameter from attractive to
repulsive interaction. We note that, as $\Lambda $ increases, the condensate
spreads, i.e., the wave function delocalizes, while its maximum decreases.
The analytical solution (\ref{ExaFi}) is less accurate at $x=0$ and $\Lambda
=-2$ (with an error less than $5\%$), in agreement with the results of Fig.~%
\ref{fig1} and the anticipated range of validity of perturbation theory.
Nevertheless, in this particular case the observed agreement with the exact
solution is quite remarkable at $x\neq 0$, and shows that the closed
analytical expressions derived here allow a fair representation of the BEC
wave function in an optical lattice. Almost no differences can be observed
on the scale of the figure between numerical and analytical solutions for $%
\Lambda =2$.

\subsection{Quenching of the non-linear interaction}

Since the repulsive atom-atom interaction and the optical lattice potential
have opposite signs, the set of parameters ($\lambda _{1D}$, $\omega $, $d$,
$sE_{R}$) can be chosen such as to minimize the effect of the non-linearity.
To study this quenching effect, in the validity range of our perturbation
approach, we consider the order parameter given by Eq.~(\ref{ExaFi}). In the
inset of the left panel of Fig.~\ref{fig5}, we monitor the dependence of $%
G(x/l_{o})$ and $F\left( x/l_{o},\alpha \right) $ on $x/l_{o}$. While the
function $G$ is universal, $F$ shows a strong dependence on the parameter $%
d/l_{o}$. Moreover, for $d/l_{o}=4$ the contribution of $F$ to the order
parameter, induced by the optical lattice, resembles the function $G$,
though with opposite sign. If we impose the condition
\begin{equation}
\Lambda G(x/l_{o})+V_{o}F\left( x/l_{o},\alpha \right) \rightarrow 0\,,
\label{cond}
\end{equation}%
then $\Phi $ can be solely described by the harmonic oscillator wave
function $\varphi _{_{0}}$. The range of values of $\Lambda /V_{o}$, $%
d/l_{o} $, and $x/l_{o}$ that satisfy (\ref{cond}) is represented by the
hatched region $\mathcal{H}$ in Fig.~\ref{fig5}. At $x=0$, Eq.~(\ref{cond})
defines the curve $\mathcal{D}$ shown in the figure. We have noted that at $%
x=0$ the functions $G$ and $F$ reach their minimum and maximum values,
respectively. In the right panel of the figure we compare the harmonic
oscillator wave function $\varphi _{_{0}}$ with the order parameter as
predicted by (\ref{ExaFi}). We chose three different points $A$, $B$, and $C$
in the parameter space of the left panel for this comparison: Indeed, $A$
does not fulfill the condition (\ref{cond}), and, consequently, $\Phi $ does
not match $\varphi _{_{0}}$. The case $C$ belongs to region $\mathcal{H}$,
but not to the curve $\mathcal{D}$. Hence, we observe a small discrepancy
between both functions, mainly at $x=0$. Case $B$, which sits right on top
of the curve $\mathcal{D}$, yields a perfect match between harmonic
oscillator wave function and order parameter. Exact numerical solution of
the GPE (\ref{GPOL}) with the parameters defined by $B$ corroborates this
result, as indicated by the open circles in the figure.

\subsection{Validity of the perturbation approach}

For a vanishing lattice, $V_{o}=0$, our closed analytical solution for the
chemical potential and its comparison with the numerical solution provides
universal criteria for its range of validity: in the interval $\left\vert
\Lambda \right\vert <2$, we derive an accuracy better than $97\%$ \cite%
{trallero,trallero2} for Eq.\ (\ref{pertmu}). For nontrivial values of the
lattice strength, two more independent parameters, $V_{o}$ and $d/l_{o}$
come into game. A necessary condition for the validity of the perturbative
results requires that the functions $\mathit{Linear}=V_{o}\exp \left(
-\alpha ^{2}\right) /2$, $\mathit{Mixed}=\Lambda V_{o}\exp \left( -\alpha
^{2}\right) \left\{ Ei(\frac{\alpha }{2})-\mathcal{C}-\ln \frac{\alpha }{2}%
\right\} /\sqrt{2\pi }$, and $\mathit{Quadratic}=V_{o}^{2}\exp \left(
-2\alpha ^{2}\right) \left\{ Chi(2\alpha )-\mathcal{C}-\ln 2\alpha \right\}
/4$, that contribute to Eq. (\ref{pertmu}), simultaneously be smaller than
one. This criterion allows to confine the range of validity of our
analytical results in a 2D map spanned by $V_{o}$ and $d/l_{o}$, as
displayed in Figs.~\ref{fig6}(a) and (b). There the $\mathit{Mixed}$
contribution to (\ref{pertmu}) is plotted for three different values of $%
\left\vert \Lambda \right\vert =0.5$, $1$, and $2$. Fig.~\ref{fig6}(a) thus
defines different $V_{o}$ vs. $d/l_{o}$ regions where Eq.~(\ref{pertmu}) can
be trusted: In region I, the quadratic term in $V_{o}$ is less important
than the linear one, while, in region II, we need to make one more
distinction, depending on the $d/l_{o}$ values. For example, if $%
d/l_{o}<(d/l_{o})_{c}$, we have that $\mathit{Quadratic}$ represents the
main contribution to $\mu $, and vice versa if $d/l_{o}>(d/l_{o})_{c}$,
where $\mathit{Linear>Quadratic}$. A similar analysis applies in regions III
and IV, with respect to the relative contributions of $\mathit{Linear}$ and $%
\mathit{Mixed}$.

Finally, let us note that experimental data as reported in \cite{Ott} on $%
^{87}$\textit{Rb} condensates were obtained for $V_{o}=275$ and $%
d/l_{o}=0.19 $. This working point is indicated by a solid circle in Fig.~%
\ref{fig6}(b), and shows that our perturbative method is applicable in the
experimental parameter range.

\section{Conclusions}

We provided closed analytical expressions for the chemical potential and the
particle density of the GPE ground state in an optical lattice. These
solutions were obtained by a perturbative expansion in the lattice strength
and in the (attractive or repulsive) atom-atom interaction, with a
well-controlled range of validity in the associated parameter space.

Interestingly, the interaction-induced non-linearity may be quenched by the
presence of the lattice. Under certain conditions (see left panel of Fig.~%
\ref{fig5}), we predict that the particle density of the repulsively
interacting system is given by $\Phi (x,t)=\exp (i\mu \lbrack \Lambda
,V_{o},d/l_{0}]t/\hslash )\varphi _{_{0}}(x/l_{0})$.

The solutions derived here can serve as a useful tool to study a weakly
interacting BEC in a not too deep 1D optical lattice. On this basis, it is
possible to develop a unified and comprehensive picture of Bogoliubov
equations, the time-dependent GPE and of collective excitations (a subject
under investigation). The model here developed can be generalized to two and
three dimensions, and also for two component BECs.

\acknowledgments C.T-G. is grateful to the Alexander von Humboldt Foundation
for financial support, and for the hospitality enjoyed during his stay at
the Max-Planck-Institut f\"{u}r Physik komplexer Systeme in Dresden. V.
L.-R. acknowledges financial support by the Brazilian agencies FAPESP and
CNPq.

\appendix

\section{Tensor $\mathbf{T}$}

Using the harmonic oscillator wave function

\begin{equation}
\varphi _{n}(x)=\left( \frac{1}{\pi ^{1/2}2^{n}n!l_{0}}\right) ^{1/2}\exp
\left( -\frac{x^{2}}{2l_{0}{}^{2}}\right) H_{n}\left( \frac{x}{l_{0}}\right)
,
\end{equation}%
with $H_{n}(z)$ the Hermite polynomials, the matrix elements $T_{plmn}$ are
defined by

\begin{eqnarray}
T_{plmn} &=&\frac{1}{\pi \sqrt{2^{n+m+l+p}n!m!l!p!}}\times  \notag \\
&&\int_{-\infty }^{\infty }\exp
(-2z^{2})H_{n}(z)H_{m}(z)H_{l}(z)H_{p}(z)dz\,.  \notag \\
&&
\end{eqnarray}
It is possible to perform the above integral, and we get \cite%
{trallero2,lord}

\begin{eqnarray}
T_{p,l,n,m} &=&\frac{(-1)^{M-m-p}2^{M-\frac{1}{2}}}{\pi \sqrt{%
2^{n+m+l+p}n!m!l!p!}}\times  \notag \\
&&\frac{\Gamma (M-l+\frac{1}{2})\Gamma (M-n+\frac{1}{2})}{\Gamma (M-n-l+%
\frac{1}{2})}\times  \notag \\
&&\text{ }_{\text{3}}F_{\text{2}}\left(
\begin{array}{cc}
-m,\text{ \ }-p, & -M+n+l+\frac{1}{2}; \\
-M+l+\frac{1}{2}, & -M+n+\frac{1}{2};%
\end{array}%
1\right) \,,  \notag \\
&&
\end{eqnarray}%
with $\Gamma (z)$ the gamma function, $_{\text{3}}F_{\text{2}}\left(
\begin{array}{ccc}
\alpha _{1}, & \alpha _{2}, & \alpha _{3}; \\
& \beta _{1}, & \beta _{2};%
\end{array}%
1\right) $ the generalized hypergeometric function \cite{Gradshteyn80}, and $%
2M=p+l+m+n$. Consequently, we arrive at the following useful relations: i) $%
T_{plmn}=0$, if $n+m+l+p$ is an odd number, ii), for $m=0$ \cite%
{Gradshteyn80},
\begin{equation}
T_{pln0}=\frac{2^{s-1}}{\pi ^{2}}\frac{\Gamma (s-l)\Gamma (s-p)\Gamma (s-n)}{%
\sqrt{2^{n+l+p}l!p!n!}}\,,
\end{equation}%
and, iii), for $p=l=n=0$,
\begin{equation}
T_{0002m}=\left( -1\right) ^{m}\frac{\sqrt{\left( 2m\right) !}}{\sqrt{2\pi }%
2^{2m}m!}\,.
\end{equation}

\section{Tensor $\mathbf{P}$}

The contribution of the optical lattice is represented by the two
dimensional matrix elements $P_{kp}$ given by

\begin{eqnarray}
P_{kp} &=&\frac{1}{\sqrt{\pi 2^{k+p}k!p!}}\int_{-\infty }^{\infty }\left[
\cos ^{2}(\frac{2\pi l_{o}}{d}z)\right. \times  \notag \\
&&\left. H_{k}(z)H_{p}(z)\exp (-z^{2})\right] dz.
\end{eqnarray}%
For $p=k+2m$ this integral is equal to \cite{Gradshteyn80}

\begin{eqnarray}
P_{kp} &=&\frac{1}{2^{m+1}}\sqrt{\frac{k!}{\left( k+2m\right) !}}\left\{
\delta _{m,0}+\right.  \notag \\
&&\left. \left( -1\right) ^{m}b^{2m}\exp \left( -b^{2}/4\right)
L_{k}^{2m}\left( \frac{b^{2}}{2}\right) \right\} \, ,  \label{P}
\end{eqnarray}%
where $b=4\pi l_{o}/d$ and $L_{k}^{t}\left( x\right) $ are the Laguerre
polynomials. The symmetry of the Hermite polynomials imposes that $P_{kp}=0$
if $p=k+2m+1$. Using (\ref{P}) the following relations hold: i) For $k=p=0$,

\begin{equation}
P_{00}=\frac{1}{2}+\frac{1}{2}\exp \left[ -\left( \frac{2\pi l_{o}}{d}%
\right) ^{2}\right]\, ,
\end{equation}%
and, ii), for $k=0$ and $p=2m$,

\begin{equation}
P_{02m}=\frac{\left( -2\right) ^{m}}{2\sqrt{(2m)!}}\left( \frac{2\pi l_{o}}{d%
}\right) ^{2m}\exp \left( -\left( \frac{2\pi l_{o}}{d}\right) ^{2}\right) \,
.
\end{equation}

\section{Series}

We can sum up the series

\begin{equation}
F1=\frac{3}{2\pi }\sum_{m=1}^{\infty }\frac{(2m-1)!}{2^{4m}(m!)^{2}}
\label{1}
\end{equation}%
noting that

\begin{equation}
\sum_{m=1}^{\infty }\frac{(2m-1)!}{2^{3m}(m!)^{2}(x^{2}+1)^{m}}=-\ln \left(
\frac{1}{2}\sqrt{\frac{2x^{2}+1}{2(x^{2}+1)}}+\frac{1}{2}\right) ^{2}\, .
\label{2}
\end{equation}%
Hence, $F1=0.033106$.

Furthermore, we have the series \cite{Gradshteyn80}

\begin{equation}
\sum_{m=1}^{\infty }\frac{1}{m2^{m}m!}\alpha ^{2m}=Ei(\frac{\alpha }{2})-%
\mathcal{C}-\ln \frac{\alpha }{2}\, ,  \label{3}
\end{equation}%
and \cite{Abramowitz}

\begin{equation}
\sum_{m=1}^{\infty }\frac{2^{2m}}{2m(2m)!}\alpha ^{4m}=Chi(2\alpha )-%
\mathcal{C}-\ln 2\alpha \, ,  \label{4}
\end{equation}%
where $Ei(x)$ is the exponential integral, $Chi(x)$ is the cosine hyperbolic
integral, and $\mathcal{C}$ is Euler's constant.

For the summation of the first part of the series in (\ref{fif}), we note
that

\begin{eqnarray}
F2(z,\alpha ) &=&\sum_{m=1}^{\infty }\frac{\left( -2\right) ^{m}\alpha ^{2m}%
}{2\sqrt{(2m)!}}\varphi _{2m}(z)  \notag \\
&=&\frac{\exp (-z^{2}/2)}{2\sqrt{l_{0}\pi ^{1/2}}}\left( \exp \alpha
^{2}\cos 2\alpha z-1\right) \, .  \label{F}
\end{eqnarray}%
The series

\begin{equation}
G1(z,\alpha )=\frac{1}{2}\sum_{m=1}^{\infty }\frac{\left( -2\right)
^{m}\alpha ^{2m}}{2m\sqrt{(2m)!}}\varphi _{2m}(z)  \label{G}
\end{equation}%
is related to the function $F2(z,\alpha )$ through the differential equation

\begin{equation}
\frac{dG1(z,\alpha )}{d\alpha }=\frac{F2(z,\alpha )}{\alpha }\,.  \label{dif}
\end{equation}%
The solution of (\ref{dif}) is given by

\begin{eqnarray}
G1(z,\alpha ) &=&\frac{1}{2\sqrt{l_{0}\pi ^{1/2}}}\exp \left( -\frac{z^{2}}{2%
}\right)  \notag \\
&&\int_{0}^{\alpha }\frac{1}{y}\left[ \exp (y^{2})\cos \left( 2yz\right) -1%
\right] dy\,.  \label{Gf}
\end{eqnarray}%
To perform the sum of the first series in (\ref{fif}), we use that
\begin{eqnarray}
F3(z,c) &=&\frac{1}{\sqrt{l_{0}\pi ^{1/2}}}\left[ c\exp (-c^{2}z^{2}+\frac{%
z^{2}}{2})-\exp \left( -\frac{z^{2}}{2}\right) \right]  \notag \\
&=&\sum_{m=1}^{\infty }\frac{\left( -1\right) ^{m}\sqrt{\left( 2m\right) !}}{%
2^{m}m!}\left( 1-\frac{1}{c^{2}}\right) ^{m}\varphi _{2m}(z)\,.  \notag \\
&&  \label{f}
\end{eqnarray}%
Now the series

\begin{equation}
g(z,c)=\sum_{m=1}^{\infty }\frac{\left( -1\right) ^{m+1}\sqrt{\left(
2m\right) !}}{m2^{m}m!}\left( 1-\frac{1}{c^{2}}\right) ^{m}\varphi _{2m}z)
\label{g}
\end{equation}%
is related to the function $F3(z,c)$ through the equation

\begin{equation}
\frac{dg}{d\beta }=\frac{2\beta }{\left( 1-\beta ^{2}\right) }F3\,,
\label{dfg}
\end{equation}%
with $\beta =1/c.$ The differential equation\ (\ref{dfg}) admits the solution

\begin{equation}
g(z,c)=\frac{2\exp \left( -\frac{z^{2}}{2}\right) }{\sqrt{l_{0}\pi ^{1/2}}}%
\int\nolimits_{1}^{1/c}\frac{\exp \left( -\frac{z^{2}}{y^{2}}\left(
1-y^{2}\right) \right) -y}{1-y^{2}}dy\, .  \label{gf}
\end{equation}


\begin{thebibliography}{99}
\bibitem{shin} Th. Anker, M. Albiez, R. Gati, S. Hunsmann, B. Eiermann, A.
Trombettoni, and M. K. Oberthaler, Phys. Rev. Lett. \textbf{94}, 020403
(2005); Y. Shin, G.-B. Jo, M. Saba, T. A. Pasquini, W. Ketterle, and D. E.
Pritchard, Phys. Rev. Lett. \textbf{95}, 170402 (2005); A. V. Ponomarev, J.
Madro\~{n}ero, A. R. Kolovsky, and A. Buchleitner, Phys. Rev. Lett. \textbf{%
96}, 050404 (2006); K. Winkler, G. Thalhammer, F. Lang, R. Grimm, J. H.
Denschlag, A. J. Daley, A. Kantian, H. P. B\"{u}chler, and P. Zoller, Nature
\textbf{441}, 853 (2006); T. Roscilde and J. I. Cirac, Phys. Rev. Lett.
\textbf{98}, 190402 (2007); J. Brand and A.R. Kolovsky, Eur. Phys. J. D
\textbf{41}, 331 (2007).

\bibitem{Groos1} E. P. Gross, Nuovo Cimento \textbf{20,} 454 (1961); L. P.
Pitaevskii, Zh. Eksp. Teor. Fiz. \textbf{40,} 646 (1961) [Sov. Phys. JETP
\textbf{13}, 451 (1961)].

\bibitem{perez} V. M. P\'{e}rez-Garc\'{\i}a, H. Michinel, J. I. Cirac, M.
Lewenstein, and P. Zoller, Phys. Rev. Lett.\emph{\ }\textbf{77}, 5320 (1996);%
\emph{\ }ibid,\emph{\ }Phys. Rev. A \textbf{56}, 1424\emph{\ }(1997); V. I.
Yukalov, E. P. Yukalova, and V. S. Bagnato, Laser Phys. \textbf{12}, 1325
(2002); ibid, Phys. Rev. A \textbf{66}, 025602 (2002);

\bibitem{Konotop} V. Konotop and P. Kevrekidis, Phys. Rev. Lett. \textbf{91,}
230402 (2003); T. Hyouguchi, R. Seto, M. Ueda, and S. Adachi, Ann. of Phys.
\textbf{312}, 177 (2004); D. Witthaut, H. J. Korsch, J. Phys. A: Math. Gen.
\textbf{39,} 14687 (2006)..

\bibitem{ana} A. M. Rey, G. Pupillo, Ch. W. Clark, and C. J. Williams, Phys.
Rew. A \textbf{72}, 033616 (2005).

\bibitem{Exp1} M. H. Anderson, J. R. Ensher, M. R. Matthews, C. E. Wieman,
E. A Cornell, Science \textbf{269,} 198 (1995).

\bibitem{Kramer} M. Kr\"{a}mer, L. Pitaevskii, and S. Stringari, Phys. Rev.
Lett. \textbf{88},\textbf{\ }180404 (2002).

\bibitem{Berg} K. Berg-S\o rensen and K. M\o lmer, Phys. Rev. A \textbf{58},
1480 (1998); S. Burger, F. S. Cataliotti, C. Fort, F. Minardi, M. Inguscio,
M. L. Chiofalo, and M. P. Tosi, Phys. Rev. Lett. \textbf{86}, 4447 (2001).

\bibitem{Pitaevskii} F. Dalfovo, S. Giorgini, L. P. Pitaevskii, and S.
Stringari, Rev. Mod. Phys. \textbf{71}, 463 (1999).

\bibitem{Pitaevskii2} L. Pezz\`{e}, L. Pitaevskii, A. Smerzi, Stringari, G.
Modugno, E. de Mirandes, F. Ferlaino, H. Ott, G. Roati, and M. Inguscio,
Phys. Rev. Lett. \textbf{93, }120401 (2004).

\bibitem{trallero} C. Trallero-Giner, J. Drake, V. L\'{o}pez-Richard, C.
Trallero-Herrero, Joseph L. Birman, Physics Letters A \textbf{354}, 115
(2006).

\bibitem{Mihling} S. G. Mikhlin and K. L. Pr\"{o}ssdorf, \textit{Approximate
Methods for Solutions of Differential and Integral Equations (}American
Elsevier Publ. Co., NY, 1967).

\bibitem{petrovskii} I. G. Petrovskii, \textit{Lectures on the Theory of
Integral Equations} (Graylock Press, Rochester, 1957).

\bibitem{trallero2} C. Trallero-Giner, J. Drake, V. L\'{o}pez-Richard, C.
Trallero-Herrero, Joseph L. Birman, Physica D, \textbf{237,} 2342 (2008).

\bibitem{Mikhlin} From general arguments it is possible to show that the
expansion (\ref{fi}) converges in the mean, i.e. $\lim\limits_{N\rightarrow
\infty }\int\limits_{0}^{\infty }\left\vert \Phi
-\sum\limits_{n=0}^{N}\varphi _{n}(x/l_{o})C_{n}(\mu )\right\vert ^{2}dx=0$
(see S. G. Mikhlin, \textit{Variational Methods in Mathematical Physics}
(Pergamon Press, 1964)). This allows to manipulate the integral in (\ref{in}%
) and the series (\ref{fi}) such as to derive the relation (\ref{Hill}).

\bibitem{Pu} H. Pu and N. P. Bigelow, Phys. Rev. Lett. \textbf{80}, 1130
(1998).

\bibitem{Ott} H. Ott, E. de Mirandes, F. Ferlaino, G. Roati, G. Modugno, and
M. Inguscio, Phys Rev. Lett. \textbf{92}, 160601 (2004).

\bibitem{lord} R. D. Lord, J. London Math. Soc. \textbf{24}, 101 (1949).

\bibitem{Gradshteyn80} I. S. Gradshteyn and I. M. Ryzhik\textit{, Tables of
Integrals, Series and Products }(Academic, NY, 1980)\textit{.}

\bibitem{Abramowitz} \textit{Handbook of Mathematical Functions,} edited by
M. Abramowitz and I. Stegun (Dover, NY, 1972)
\end{thebibliography}
\end{document}